\documentstyle[12pt,epsf]{article}
\newcommand{\al}{\alpha}

\newcommand{\dl}{\delta}
\newcommand{\Dl}{\Delta}

\newcommand{\veps}{\varepsilon}

\newcommand{\sg}{\sigma}

\newcommand{\Ss}[1]{\mbox{$\cal #1$}}
\newcommand{\beq}{\begin{equation}}
\newcommand{\bea}{\begin{eqnarray}}
\newcommand{\beas}{\begin{eqnarray*}}
\newcommand{\eeq}{\end{equation}}
\newcommand{\eea}{\end{eqnarray}}
\newcommand{\eeas}{\end{eqnarray*}}
\newcommand{\ebar}{\bar{e}}

\newcommand{\psibar}{\bar{\psi}}
\newcommand{\sgbar}{\bar{\sg}}
\newcommand{\Ahat}{\hat{A}}
\newcommand{\What}{\hat{W}}

\newcommand{\Atil}{\tilde{A}}

\newcommand{\Jvec}{\vec{J}}

\newcommand{\Kkw}{\vec{K}^2}

\newcommand{\Leen}{{\vec{L}_1}}
\newcommand{\Leenkw}{\vec{L}_1^2}
\newcommand{\Lkw}{\vec{L}^2}
\newcommand{\Ltw}{{\vec{L}_2}}
\newcommand{\Ltwkw}{\vec{L}_2^2}

\newcommand{\Svec}{\vec{S}}
\newcommand{\Tvec}{\vec{T}}
\newcommand{\half}{\sfrac{1}{2}}

\newcommand{\thalf}{\sfrac{3}{2}}
\newcommand{\fhalf}{\sfrac{5}{2}}

\newcommand{\nhalf}{\sfrac{9}{2}}

\newcommand{\ad}{\mbox{\,ad\,}}

\newcommand{\spur}[1]{{\,\mbox{sp}_{#1}}}
\def\tr{\mathop{\rm tr}\nolimits}
\newcommand{\trees}[2]{\,\mbox{tr}_{#1} \left( #2 \right)}

\newcommand{\form}{{\small FORM}}

\newcommand{\mathem}{{\small MATHEMATICA}}
\newcommand{\cl}{{\mbox{cl}}}

\newcommand{\drs}{{three-sphere}}

\newcommand{\eff}{{\mbox{eff}}}

\newcommand{\EUC}{{\mbox{\scriptsize EUCL}}}
\newcommand{\id}{{\unitmatrix{.6}}}
\newcommand{\ie}{{i.e.\ }}
\newcommand{\Next}{\nonumber \\ }
\newcommand{\pr}{\partial}
\newcommand{\Sdr}{{$S^3$}}
\newcommand{\SUtw}{{$SU(2)$}}
\newcommand{\sutw}{{$su(2)$}}
\newcommand{\un}{\underline}
\newcommand{\unitmatrix}[1]{
        \thinlines
	\setlength{\unitlength}{#1mm}
	\begin{picture}(2.2,5)
	\put(0.7,0){\line(0,1){4.2}}
	\put(1.5,0){\line(0,1){5}}
	\bezier{15}(0.0,3.5)(0.75,4.25)(1.5,5.0)
	\put(0  ,0){\line(1,0){2.2}}
	\end{picture} }

\newcommand{\Order}[1]{\Ss{O}\left(#1\right)}
\newcommand{\real}{\relax{\rm I\kern-.18em R}}

\newcommand{\sfrac}[2]{\mbox{\large $\frac{#1}{#2}$}}

\begin{document}
\vskip-1cm
\hfill INLO-PUB-3/96
\vskip5mm
\vskip10mm
\begin{center}
{\LARGE{\bf{
  \un{One-loop effective action for}
  \un{\SUtw\ gauge theory on \Sdr}
}}}\\
\vspace*{5mm}
\vspace*{1cm}{\large Bas van den Heuvel\footnote{e-mail:
bas@lorentz.LeidenUniv.nl}} \\
\vspace*{1cm}
Instituut-Lorentz for Theoretical Physics,\\
University of Leiden, PO Box 9506,\\
NL-2300 RA Leiden, The Netherlands.\\ 
\end{center}
\vspace*{5mm}{\narrower\narrower{\noindent
\underline{Abstract:}  We consider the effective theory for the low-energy
modes of \SUtw\ gauge theory on the three-sphere. By
explicitely integrating out the high-energy modes, 
the one-loop correction to the hamiltonian for this problem is obtained.
We calculate the influence of this correction on the glueball spectrum.
}\par}

\section{Introduction} 
Non-perturbative effects in gauge theories can be studied
using finite volumes~\cite{baa2,baa1,baa6}. 
In a small volume asymptotic freedom implies that the coupling
constant is small: this means that we can use standard perturbation theory. 
By increasing the volume, 
we can study the onset of non-perturbative phenomena.
The effects that we want to study are 
related to the multiple vacuum structure of the theory.
For increasing volume, the wave functional starts to spread out
over the configuration space in those directions where
the potential energy is lowest, \ie in the direction of the
low-energy modes of the gauge field. In particular, it will
flow over the instanton barrier that connects gauge copies of
the vacuum.
\par
As long as the non-perturbative effects manifest themselves
appreciably only in a small number of low-lying energy modes, this can be
described adequately using a hamiltonian formulation.
It is hence our strategy to split up the gauge field in orthogonal modes
and to reduce the dynamics of this infinite number of degrees of freedom
to a quantum mechanical problem with a finite number of modes.
\par
In our approach, we impose the Coulomb gauge by restricting the 
gauge fields to a so-called fundamental domain~\cite{sem,zwan}.
The spreading out over configuration space means that
the wave functional will become sensitive to the
boundary conditions that have to be imposed on the boundary
of this fundamental domain. In these boundary conditions
the dependence on the $\theta$-angle will show up.
\par
For more details on this method the reader is referred 
to~\cite{heuf1}, where we used the lowest order 
effective hamiltonian in a variational calculation of the spectrum.
In the present letter we perform the one-loop calculation, 
that is, we integrate out the high-energy modes in the path integral 
to obtain the correction to the lowest order hamiltonian. 
We will subsequently use this new hamiltonian to find the glueball spectrum.

\section{The effective theory}

We will briefly review the technical set up for the analysis on the
\drs. For details and more motivation, we again refer to~\cite{heuf1}
and references therein. Let $n_\mu$ be the normal vector on the 
\drs.
We define two orthonormal framings on \Sdr\ by
\beq
  e^\al_\mu = \eta^\al_{\mu \nu} n_\nu , \hspace{1.5cm}
  \bar{e}^\al_\mu = \bar{\eta}^\al_{\mu \nu} n_\nu,
\eeq
where we used the 't Hooft $\eta$ symbols~\cite{tho}.
These $\eta$ symbols occur in the multiplication rules for
the unit quaternions:
\beq
 \sg_\mu \sgbar_\nu = \eta^\al_{\mu \nu} \sg_\al, \hspace{1.5cm}
  \sgbar_\mu \sg_\nu = \bar{\eta}^\al_{\mu \nu} \sg_\al,
\eeq
where $\sg_\mu$ and their conjugates 
$\sgbar_\mu = \sg^\dagger_\mu$ are defined by
\beq
  \sg_\mu = ( \id ,i \vec{\tau}), \hspace{1.5cm} 
  \bar{\sg}_\mu = ( \id,- i \vec{\tau}).
\eeq
We choose to write a gauge field on \Sdr\ ($n_\mu A_\mu = 0$) 
with respect to the framing $e^i_\mu$:
\beq
   A_\mu= A_i e_\mu^i = A^a_i e_\mu^i \frac{\sg_a}{2}
\eeq 
We introduce a number of \sutw\ angular momentum operators.
We define $L^i \equiv L_1^i = \sfrac{i}{2}\, e^i_\mu \pr_\mu$ and
$L_2^i = \sfrac{i}{2} \, \ebar^i_\mu \pr_\mu$. These operators generate
the \SUtw\ $\times$ \SUtw\ symmetry of \Sdr\ and satisfy
$\Leenkw = \Ltwkw$. We introduce a spin operator
$\Svec$ by $ (S^a A)_i = - i \veps_{a i j} A_j$ and
an isospin operator $\Tvec$ by $T^a = \ad(\tau_a/2)$.
We also define $K_i = L_i + S_i$ and $J_i = K_i + T_i$.
\par
To isolate the lowest energy levels, we write
\beq
  V_\cl(A) =  - \frac{1}{2 \pi^2} \int_{S^3} \frac{1}{2} \tr(F_{ij}^2) 
  = - \frac{1}{2 \pi^2} \int_{S^3} \tr(A_i \Ss{M}_{ij} A_j) + \Order{A^3}, \label{fluctdef}
\eeq
where the quadratic fluctuation operator \Ss{M} can be rewritten as
\beq
\Ss{M}_{ij} = \left( \Kkw - \Lkw \right)^2_{ij}
\eeq
The zero-modes of \Ss{M} correspond to pure-gauge modes of the gauge field.
The $18$ dimensional space $B(c,d)$ given by
\beq
    B_\mu(c,d) = \left(c^a_i  e^i_\mu+d^a_j\bar{e}^j_\mu \right) 
  \frac{\sg_a}{2} 
= \left(c^a_i + d^a_j V^j_i \right) e^i_\mu \frac{\sg_a}{2}
\label{Bcddef}
\eeq
is the eigenspace of \Ss{M} corresponding to its lowest positive 
eigenvalue 4, whereas the next eigenvalue is 9.
The tunnelling path
is $c^a_i = -u \, \dl^a_i, \, d^a_i = 0$ with $u$ running from $0$ to $2$.
For $u=1$ it passes through the sphaleron, which is a saddle point of the
energy functional.
The energy functional for these 18 modes is given by
\beq
  V_\cl(c,d) \equiv - \frac{1}{2 \pi^2} \int_{S^3} \frac{1}{2} \tr(F_{ij}^2)
  = V_\cl(c) + V_\cl(d) + \sfrac{1}{3}
   ( \tr(X) \tr(Y) - \tr(X Y) ), \label{pot}
\eeq
\beq
  V_\cl(c) =  2 \tr(X) + 6 \det c + 
  \sfrac{1}{4} \left( \mbox{tr}^2(X) - \tr(X^2) \right),
\eeq
with the symmetric matrices $X$ and $Y$ given by 
$X = c \, c^T$ and $Y = d \, d^T$.
The lowest order hamiltonian for these modes is
\beq
  R~\Ss{H}(c,d) = -\frac{f}{2} \left( 
  \frac{\pr^2}{\pr c^a_i \pr c^a_i} + \frac{\pr^2}{\pr d^a_i \pr d^a_i} 
   \right) + \frac{1}{f} V_\cl(c,d)\label{hamdef}
\eeq
with $f = \frac{g_0^2}{2 \pi^2}$, and $R$ the reinstated radius of the
sphere.
\par

\section{Gauge fixing}

We will impose the background gauge condition on the high-energy modes.
Consider a general gauge field on \Sdr:
\beq
  A_\mu = \left( A_0, A_i \right),
\eeq
where $A_0$ is the time component of the gauge field and $A_i$
are the space components with respect to the framing $e^i_\mu$.
We will now project out the background field $B(c,d)$. Let $P_S$ be 
the projector on the constant scalar modes, and let $P_V = P_c + P_d$
be the projector on the $(c,d)$-space.
We define the background field $B$ and the quantum field $Q$ by
respectively
\bea
  B_\mu &=& ( P A )_\mu = \left( P_S A_0, (P_V A)_i \right), \\
  Q_\mu &=&  A_\mu - B_\mu.
\eea
We define the gauge fixing function $\chi$ by
\beq
  \chi = \left (1 - P_S \right)D_\mu(PA) A_\mu + P_S A_0.
\eeq
We use $\chi$ to impose the background gauge condition: $\chi = 0$ 
is equivalent to $B_0 = 0$ and $D_\mu(B) Q_\mu  = 0$.
We perform the standard manipulations with the partition function:
after introducing Faddeev-Popov ghosts and expanding the
classical action up to second order in $Q$ we obtain
\bea
Z  
&=&
\int DB_k\, D'Q_\mu\, D'\psi\, D'\psibar  \exp\left[ \frac{1}{g_0^2} 
 \int \tr \left( \psibar \left\{ - D_\mu(B) (1 - P) D_\mu(B) \right\} \psi 
  \right. \right. \Next
&& + \left. \left.  \half F_{\mu \nu}^2(B) - 
  2 (D_\mu F_{\mu \nu})(B) Q_\nu + 
  Q_\mu W_{\mu \nu}(B) Q_\nu \right) \vphantom{\frac{1}{g_0^2}} \right],
\eea
with
\beq
\left\{ \begin{array}{ll}
W_{0 0} &= -D_\rho^2(B) \\
W_{0 i} &= - W_{i 0} = - 2 \ad (\dot{B}_i) \\
W_{i j} &= -2 \ad(F_{i j}(B)) - (D_\rho^2(B))_{i j}
+ 2 \dl_{i j}
\end{array}
\right.
\eeq
Remember that the covariant derivative $D_i(B)$ acting on vectors (or tensors)
gives extra connection terms (due to $S^3$ being a curved manifold), e.g.
\bea
  (D_i(B) C)_j &=& \pr_i C_j + [B_i, C_j] - \veps_{i j k} C_k \\ 
 &=& \left(- 2 i L_i + i B_i^a T^a - i S_i \right)_{j k} C_k
\eea
The primed integration means that we have excluded 
the $(c,d)$-modes from the integration over the vector field $Q_k$, and
the constant modes
from the integration over the scalar fields $\psi$, $\psibar$ and $Q_0$. 
\par
The action contains a term $J_\nu Q_\nu$ with 
$J_\nu = (D_\mu F_{\mu \nu})(B)$. 
Since $B$ need not satisfy the equations of motion,
this term does not vanish.
When expanding the path integral in Feynman diagrams, this term
will give rise to extra diagrams, where $J$ acts as a source.
It can be shown that $J$ will only contribute to
terms in the effective lagrangian that we will consider to
give only small corrections: 
they are at least of the order $c^2 d^4$ or $c^4 d^2$ and we will
ignore them.
\par
Dropping the term linear in $Q_\mu$, we obtain from $Z$
the effective action
\bea
&& S^\eff_\EUC[B] \Next
&=& S^\cl_\EUC[B] 
- \ln {\det}' \left( - D_\mu(B) (1 - P) D_\mu(B) \right)
+ \half \ln {\det}' \left( W_{\mu \nu} \right) 
\label{Seff}
\eea
 
\section{The effective potential}
As the one-loop computation is the central result for this paper 
we present here a few of the details that are crucial to understand 
why we were able to complete this calculation.
For computing the effective potential, $B$ is considered to be 
independent of time.
This effective potential is determined up to fourth order in
the fields $c$ and $d$ and up to sixth order in the
tunnelling parameter $u$, where we expect the physics 
to be most sensitive to the precise shape of the potential.
We use $A(B)$ to denote any of the operators occuring in~(\ref{Seff}).
They are of the form
\beq
  A(B) = -\pr_0^2 - \pr_\veps^2 + \Atil(B).
\eeq
In order to neatly perform the dimensional regularisation,
we introduced a laplacian term for an $\veps$-dimensional
torus of size $L$ that we attached to our space \Sdr~\cite{lus1}.
The scale $L$ should of course be chosen proportional to the
radius $R$ of the \drs. 
The assumption $[\pr_0, \Atil(B)] = 0$ allows us to write
\beq
  \ln \det (A)  =
-\frac{T}{2 \sqrt{\pi}} \left( \frac{L}{2 \sqrt{\pi}} \right)^\veps
\int_0^\infty ds\, s^{-3/2 - \veps/2} \tr \left( e^{-s \Atil} \right),
\label{lndetA2}
\eeq
where we took the time periodic with period $T$.
The operator $\Atil$ can be expressed in terms of the
angular momentum operators defined above, the functions $V^j_i$ 
and the constants $c^a_i$ and $d^a_i$.
To take the trace, we need a basis of functions. For the scalar
operators (the ghost operator and $W_{0 0}$), 
we can use $ |l\, m_L\, m_R \rangle \otimes |1\, m_t\rangle$
or equivalently $| l\,m_R; j\,m_j \rangle$, where $m_L$, $m_R$ and
$m_j$ correspond to the z-components of $\Leen$, $\Ltw$ and $\Jvec$ respectively.
Here $l=0,\half,1,\ldots$ and $j = |l-1|, \ldots, l+1$.
For the vector operator $W_{i j}$, we can use
$ |l\, m_R ; k\, m_k \rangle \otimes |1\, m_t\rangle$ or
$| l\,m_R;k; j\,m_j \rangle$, where the bounds on
the various quantum numbers are obvious. 
Note that the $c$ and $d$ modes correspond to the vector modes with 
$(l,k) = (0,1)$ and $(l,k) = (1,0)$ respectively.
For the scalar operators, the trace
must not be taken over the $l=0$ functions, whereas 
for the vector operator
the trace must not include the $c$ and $d$ modes.
Note that for the case of the ghost operator, the operator $(1-P)$
(see eq.~(\ref{Seff})) 
causes some intermediate vector modes to be projected to zero.
\par
Using the appropriate basis,
we can calculate the determinants exactly for the vacuum $u=0$ ($B=0$),
and for the sphaleron configuration $u=1$ ($c_i^a=-\dl_i^a$ and $d=0$). 
The final summation
over $l$ is expressed using the  function $\zeta(s,a)$, 
which is defined by
\beq
\zeta(s,a) = \sum^\infty_{k=2} \frac{1}{(k^2 + a)^s}, \quad 
\mbox{Re}(s) > \half,
\label{zetadef}
\eeq
After analytic continuation we have
\beq
\zeta(s,a) = \sum^\infty_{m=0} \frac{(s)_m}{m !}\,  
  (-a)^m\, (\zeta_R(2 s + 2 m) - 1 ),\quad s \neq \half,-\half,\ldots,
\label{zetaexp}
\eeq
where $\zeta_R$ denotes the Riemann $\zeta$-function and
$(s)_m$ is Pochhammer's symbol.
If $s$ approaches one of the poles, this expansion can be
used to split off the divergent term: the remainder of the
series is denoted with $\zeta_F(s,a)$. We find for the
effective potentials
\bea
\Ss{V}^{(1)}(B=0) &=& -18 + 3\,\zeta_R(-3) - 3\,\zeta_R(-1), \\
 \Ss{V}^{(1)}(\mbox{Sph})  &=&
  -1 - 3\,{\sqrt{2}} - 12 \sqrt{3} 
   + \nhalf \sqrt{6} - \fhalf \sqrt{10}
   + \sfrac{11}{4\,\veps} - 
   \sfrac{11}{8}\,\gamma + \sfrac{11}{4} \log (2) 
\Next && 
+  \sfrac{11}{4}\,\log \left(\sfrac{L}{2 \sqrt{\pi}} \right) 
  +{\zeta_F}(-\thalf,-3) + {\zeta_F}(-\thalf,-2) + 
   {\zeta_F}(-\thalf,1) 
\Next && 
  + 3\,{\zeta_F}(-\half,-3)
  +  {\zeta_F}(-\half,-2) - 5\,{\zeta_F}(-\half,1)
\eea
The pole term $\sfrac{11}{4\,\veps}$ is absorbed through
the usual renormalisation of the coupling constant
\beq
  \frac{1}{g_R^2} =  \frac{1}{g_0^2} + \frac{11}{12 \pi^2 \veps}
  +  \mbox{finite renormalisation}
  \label{infren}
\eeq
For a general configuration along the tunneling path, we make an
expansion of the eigenvalues of the operators in $u$.
Although it is still possible to calculate the spectra
exactly, a polynomial form of the effective potential is much
more useful in the variational method that we will use.
We used Bloch perturbation theory~\cite{blo} to obtain
the expansions of the eigenvalues of $W_{i j}$.
Results up to tenth order in $u$ were obtained, where we used both
\mathem\ and \form\ in the calculations.
\par
As can be seen from fig.~1, the expansions do not converge
to the exact result at $u=1$. This should come as no big
surprise, since we have no reason to expect the radius of convergence
of the expansion to be as large as one. 
To find the effective potential for larger $u$, we write $u=1+a$ and
make a similar expansion of the effective potential around
the sphaleron. 
Using the fourth order expansion in $u$ and the first order expansion
in $a$ (\ie the value and the slope of the potential 
at the sphaleron), we can construct a polynomial in $u$ of degree six
that is a good approximation to the effective potential.
\par
We now turn to $B = B(c)$ (we still keep $d=0$). 
The perturbative evaluation
of the individual eigenvalues is no longer possible, but we can
use the following technique.
Suppose we have $\Atil = F + \Ahat$ with $F = \Atil(0)$
such that $[F,\Ahat] = 0$.
This allows us to substitute in~(\ref{lndetA2})
\beq
  \tr \left( e^{-s \Atil} \right) =
  \tr \left( e^{-s F} e^{-s \Ahat} \right) =
  \sum_i e^{-s F_i} \sum_{n=0}^\infty 
    \frac{(-s)^n}{n!} \trees{i}{ \Ahat^n}.
\label{lndetA3}
\eeq
The sum over $i$ is a sum over the eigenspaces of $F$, $F_i$ is the
corresponding eigenvalue and
$\trees{i}{}$ denotes a trace within the eigenspace.
For both the scalar operators we have $F = 4 \Lkw$ and $[\Lkw,\Ahat(c)] = 0$.
The remaining problem
of calculating $\trees{l}{\Ahat^n(c)}$ then reduces to calculating 
traces of the form
\beq
  \trees{l} {L_{i_1} \ldots L_{i_n} T^{a_1} \ldots T^{a_m} }
\eeq
which can be done relatively easy.
For the vector operator we have $F = 2 \Lkw + 2 \Kkw$ and
$[\Kkw,\What(c)] \neq 0$.
The trace in~(\ref{lndetA3}) can however still be written as
a sum of traces in the different eigenspaces of $F$:
\bea
&&  \tr \left( e^{-s \Atil} \right)
\Next &=&  
   \sum_i e^{-s F_i} \trees{i}{ e^{-s (F - F_i + \Ahat)}}
\Next &=&  
   \sum_i e^{-s F_i} \sum_{n=0}^\infty \frac{(-s)^n}{n!} 
  \trees{i} { \left\{(F - F_i) + \Ahat\right\}^n }
\Next &=& 
   \sum_m \spur{m}
\label{lndetA4}
\eea
Here $\spur{m}$ is the contribution of order $m$ in $\Ahat$. 
Let $P_i$ denote the projector
on the eigenspace of $F_i$, and let \Ss{T} be given by
\bea
  \Ss{T}_i &=& \trees{i}{\Ahat} = \tr (P_i \Ahat), \\
  \Ss{T}_{ i j \cdots} &=&  \tr (P_i \Ahat P_j \Ahat \cdots).
\eea
For a given value of $m$ we can perform the combinatorics 
to write $\spur{m}$ in terms
of the \Ss{T} functions. With $\Dl_{j i} = F_j - F_i$ we find
\bea
  \spur{1} &=& \sum_i e^{-s F_i} (-s) \Ss{T}_i, \\
  \spur{2}  &=&
 \sum_i e^{-s F_i} \sum_{n=2}^\infty \frac{(-s)^n}{n!} 
  \trees{i} { \Ahat (F - F_i)^{n-2} \Ahat } 
\Next &=& 
 \sum_{i j} e^{-s F_i} \sum_{n=2}^\infty \frac{(-s)^n}{n!} 
  \Ss{T}_{i j} \Dl_{j i}^{n-2}
\Next &=& 
  \sum_i \Ss{T}_{i i} e^{-s F_i} \frac{s^2}{2} 
  + \sum_{i \neq j}  \Ss{T}_{i j} e^{-s F_i} \frac{s}{\Dl_{j i}},
\\  \spur{3} &=&
  \sum_{i j k} \Ss{T}_{i j k} e^{-s F_i}  
  \sum_{n=3}^\infty \frac{(-s)^n}{n!} 
  \sum_{m_1 = 0}^{n-3} \Dl_{j i}^{m_1} \Dl_{k i}^{n - 3 -m_1},
\label{spexp}
\eea
etc.
\par
Starting from one eigenspace $i_m$, the number of intermediate states
that can be reached is finite. This allows us to extract one overal summation,
and to perform the remaining finite sums.
For the case at hand, we extract the summation over $l$ and are left
with an $l$-dependent expression in which the \Ss{T} functions have
the form
\beq
 \Ss{T}_{k_1 k_2 \ldots} = \trees{l}{P^K_{k_1} \What P^K_{k_2} \What \cdots}.
\eeq
Since the intermediate modes in the \Ss{T} functions 
can only have $k = l-1,\ldots, l+1$, we can write
\beq
P^K_k = a_0(k) + a_1(k) \Kkw + a_2(k) (\Kkw)^2,
\eeq
for certain values of the coefficients. Note that a \Ss{T} function
with an intermediate value $k=0$ should be discarded, since
this corresponds to a $d$ mode.
The combinatorics for $\spur{m}$, the finite sums, 
the evaluation of the \Ss{T} functions and
the final summation over $l$ were all done in \form. 
\par
For general $B(c,d)$, not even $\Lkw$ commutes with the
various operators. The methods described above are however
sufficient. The \Ss{T} traces that we have to calculate
now also contain the operators $\Ltw$ and explicit $V$ functions, as well
as projectors on different $\Lkw$ intermediate levels.
Using the $c \leftrightarrow d$ symmetry it is however
only necessary to obtain the $c^2 d^2$ terms.
Since the precise form of the coefficients is not
very illuminating, we postpone writing down the effective potential
until we have performed the renormalisation.

\section{The renormalisation}

To obtain the one-loop contribution to the operator $\dot{B}^2$,
we perform the usual expansion of the path integral in
Feynman diagrams. The subtleties related to the summation
over the space-momenta were dealt with in the previous section.
The diagrams needed are depicted in fig.~2, where
the particle in the loop is respectively a ghost, a scalar $Q_0$
or a vector particle.
The insertions in the diagrams correspond to the operator $\Ahat$
defined above. 
The propagators in momentum space are given by
\beq
  \frac{-1}{ k_0^2 + k_\veps^2 + 4 l (l+1) }
\eeq
for scalar particles, and
\beq
  \frac{-1}{ k_0^2 + k_\veps^2 + 2 l (l+1) + 2 k (k+1) }
\eeq
for vector particles. Here $k_0$ is the time component of the momentum
and $k_\veps$ is a momentum related to the $\veps$-dimensional torus.
\par
The $\dot{B}^2$ term comes from the
diagrams with two insertions. If the two time-momenta in
these diagrams are denoted by $p$ and $p+q$, we first perform
the integration over $p$, and then expand the result in $q^2$.
Using partial integration, these powers of $q^2$ can be
transformed in time-derivatives acting on $\Ahat$ and hence on $B$.
There is also a diagram with an explicit dependence on $\dot{B}$.
It is the diagram with two insertions of the operator $W_{0 i}$,
one $Q_0$ and one $Q_i$ propagator.
\par
Adding up the different contributions, we obtain the
one-loop contribution to the kinetic term 
 $\dot{c}^a_i \dot{c}^a_i + \dot{d}^a_i \dot{d}^a_i$
in the lagrangian. Demanding the renormalised kinetic part to look 
just like the classical term gives us the finite part of the
renormalisation~(\ref{infren}):
\beq
   \frac{1}{g^2_R} = \frac{1}{g^2_0} + 
    {\frac{11}{12 \pi^2  \veps}} +
     \frac{11}{12 \pi^2 }\,\log (\sfrac{L}{2 \, \sqrt{\pi}}) + \kappa_0,
\label{renorm}
\eeq
where $\kappa_0$ can be found in table~\ref{kappadat}.
This renormalisation scheme can easily be related to other 
schemes like the $\overline{MS}$ scheme.
\par
The finite, renormalised effective potential becomes
\beq
  \Ss{V}_\eff = \frac{2 \pi^2}{g_R^2} \, V_\cl(c,d) + V_\eff^{(1)},
\eeq
with
\bea
V_\eff^{(1)}(c,d) &=&   
V_\eff^{(1)}(c) + V_\eff^{(1)}(d) + 
  \kappa_7 \, {\tr}(X)\,{\tr}(Y) +  \kappa_8 \, {\tr}(X\,Y), \Next
V_\eff^{(1)}(c) &=&   
   \kappa_1 \, {\tr}(X)  + 
   \kappa_2 \, {\det}(c) +
   \kappa_3 \, {\mbox{tr}^2}(X) +
   \kappa_4 \, {\tr}({X^2}) 
\Next &&  +
   \kappa_5 \, {\det}(c)\,{\tr}(X) +
   \kappa_6 \, {\mbox{tr}^3}(X),
\eea
with the numerical values for $\kappa_i$ in table~\ref{kappadat}.
\begin{table}[t]
\centering
\begin{tabular}{|l@{ = }r@{.}l|} \hline
$\kappa_0$ & $ 0$&$0566264741439181  $\\ \hline
$\kappa_1$ & $-0$&$2453459985179565 $\\
$\kappa_2$ & $ 3$&$66869179814223   $\\
$\kappa_3$ & $ 0$&$500703203096610  $\\
$\kappa_4$ & $-0$&$839359633413003  $\\
$\kappa_5$ & $-0$&$849965412245339  $\\
$\kappa_6$ & $-0$&$06550330854836428$\\
$\kappa_7$ & $-0$&$3617122159967145 $\\
$\kappa_8$ & $-2$&$295356861354712  $\\ \hline
\end{tabular}
\caption{The numerical values of the coefficients.}
\label{kappadat}
\end{table}
Note that the $u^5$ term in the effective
potential along the tunnelling path uniquely determines the
coefficient of the $\tr(X) \det(c)$ term.
The $u^6$ term can be obtained from
combinations of the three independent invariants $\mbox{tr}^3(X)$, 
$\tr(X) \, \tr(X^2)$ and $\tr(X^3)$.
We choose to replace the $u^6$ term by $\mbox{tr}^3(X)$,
which is the simplest of these from the viewpoint of the
variational calculation.

\section{Variational Results and Conclusions}

We calculated the effect of the high-energy modes on the dynamics of
the low-energy modes. It resulted in a renormalisation of the
coupling constant and a correction to the potential in the
effective hamiltonian.
\par
With the obtained hamiltonian, we can repeat
the variational approximation~\cite{heuf1} of the spectrum.
The results remain qualitatively the same: the lowest-lying 
scalar ($j = 0$) and tensor ($j=2$) levels can be found in the
same sectors as before, although the values of the energy
have changed. Results for the lowest glueball masses can
be found in fig.~3. Around $f=0.25$ the 
mass ratio $m_{2^+}/m_{0^+}$ is roughly $1.5$, which 
compares nicely with the lattice results~\cite{mic}.
\par
Just as for the lowest-order hamiltonian, 
we used Temple's inequality~\cite{ree} to
obtain lower bounds for the energy levels.
This convinced us as before that we obtained accurate results.
However, due to the more complicated structure of the
potential, a larger number of basis vectors is required.
\par
The onset of the influence of the boundary can be seen at
$f= 0.2$. 
One of the issues raised in~\cite{heuf1} was the level of
localisation of the wave function around the sphaleron.
This is related to the question whether the assumption is true that
only the boundary conditions at and near the sphalerons are felt.
We argued that this was determined by the rise of the potential
in the transverse directions. The one-loop correction to the
$\tr(Y)$ term in the potential at the $c$-sphaleron, which can be
expressed in $\kappa_1$, $\kappa_7$ and $\kappa_8$, is such that
it results in a lesser degree of localisation. 
Beyond $f = 0.3$ the approximation breaks
down as can be seen for instance by the crossing of the
scalar and tensor glueball. 
A fuller study into this localisation, 
as well as more details and results will be presented 
in the near future.

\section{Acknowledgment}

The author wishes to thank Pierre van Baal for many
helpful discussions on the subject.

\newpage
\begin{center}
\Large{\bf Figure captions}
\end{center}
\vskip1cm
{\narrower\narrower{\noindent
Figure 1: 
Expansion of $\Ss{V}^{(1)}(u)$ in the tunneling parameter $u$.
We dropped the $\veps$ and $\log (L)$ dependent parts. We have
drawn the expansion up to order $u^4$, $u^6$, $u^8$ and $u^{10}$.
Longer dashes correspond to higher order expansions. The horizontal
line at $u=1$ denotes the exact result at the sphaleron.
}\par}
\vskip1cm
{\narrower\narrower{\noindent
Figure 2: 
Topology of the different Feynman diagrams.
}\par}
\vskip1cm
{\narrower\narrower{\noindent
Figure 3: Glueball masses for $\theta = 0$ as a function of 
the coupling constant. 
The lower and upper drawn curves are the masses of resp. the 
first scalar ($0^+$) and
tensor ($2^+$) glueball. 
The dotted lines denote the perturbative results.
}\par}
\vskip1cm

\newpage

\begin{figure}[t]
\centering
\epsfxsize=0.8\textwidth
\leavevmode
\epsffile{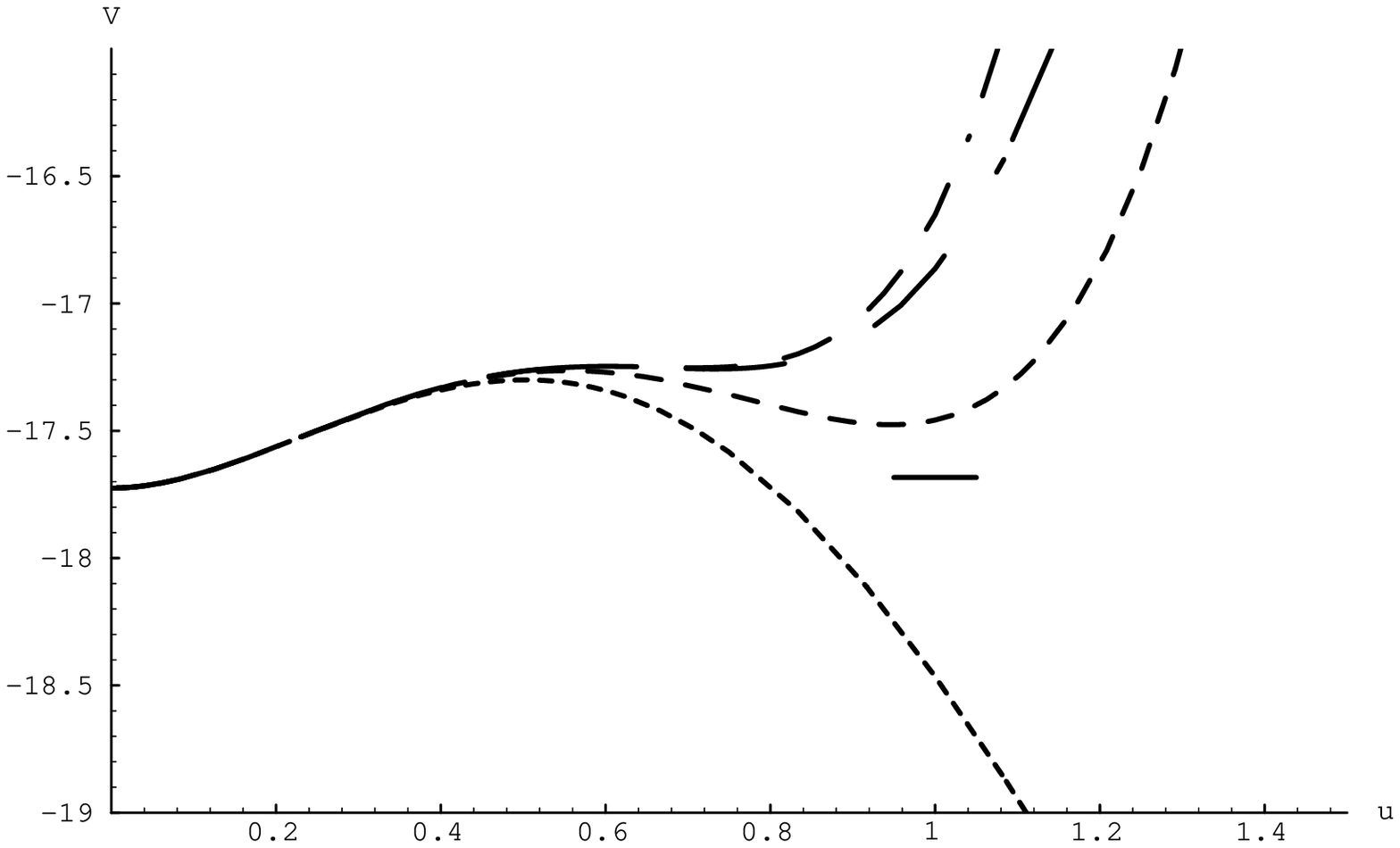}
\end{figure}

\begin{figure}[t]
\centering
\epsfxsize=.4\textwidth
\leavevmode
\epsffile{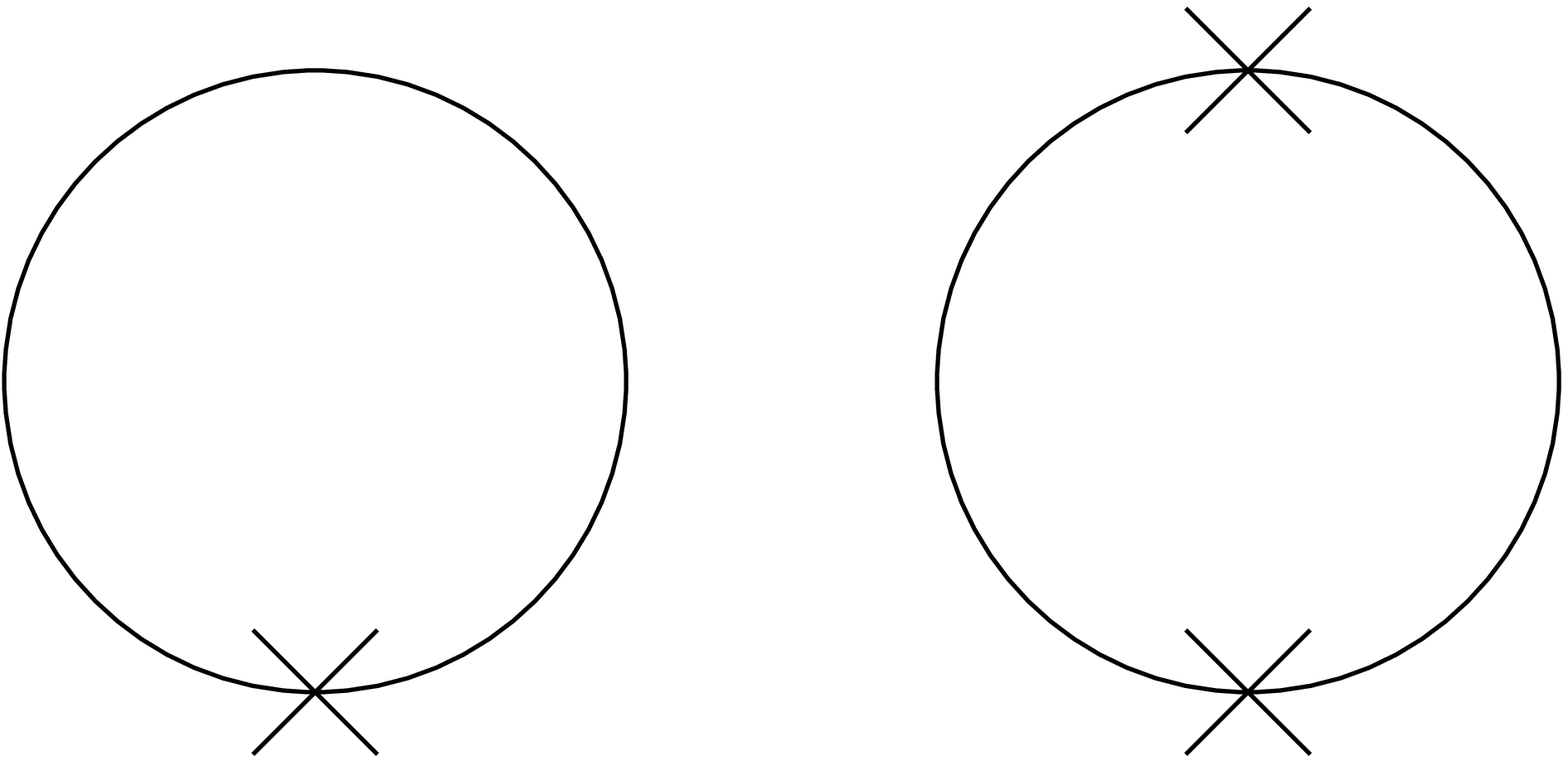}
\end{figure}

\begin{figure}[t]
\epsfxsize=152mm
\epsffile{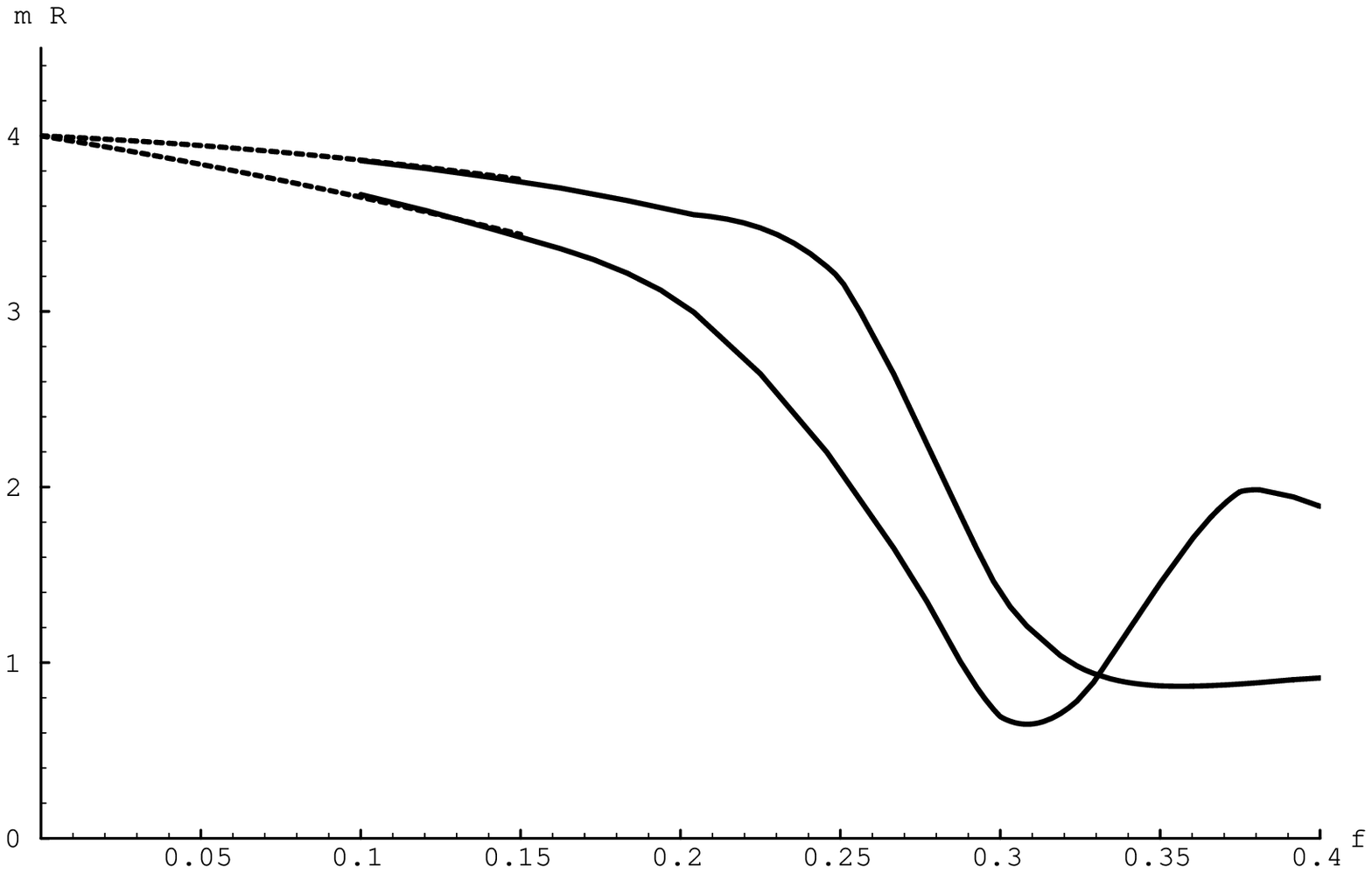}
\end{figure}

\end{document}